\documentclass[letter]{aa}
\usepackage{natbib}
\usepackage[binary-units=true]{siunitx}
\bibpunct{(}{)}{;}{a}{}{,}

\usepackage{graphicx}
\usepackage{txfonts}

\usepackage{soul}
\newcommand{\hrieuv}{HRI\textsubscript{EUV}\xspace}
\newcommand{\hrilya}{HRI\textsubscript{Lya}\xspace}

\begin{document}

   \title{Extreme UV quiet Sun brightenings observed by Solar Orbiter/EUI}


   \author{D. Berghmans\inst{\ref{i:rob}}\fnmsep\thanks{Corresponding author: David Berghmans \email{david.berghmans@oma.be}}
        \and
        F. Auch\`ere\inst{\ref{i:ias}}
        \and
        D. M. Long\inst{\ref{i:mssl}}
        \and
        E. Soubri\'e\inst{\ref{i:ias},\ref{i:iaccc}}
        \and
        M. Mierla \inst{\ref{i:rob}, \ref{i:igar}}
        \and
        A.~N.~Zhukov\inst{\ref{i:rob}, \ref{i:sinp}}
        \and
        U. Sch\"uhle\inst{\ref{i:mps}}
        \and
        P. Antolin\inst{\ref{i:north}}
        \and
        L. Harra\inst{\ref{i:pmod},\ref{i:eth}}
        \and
        S. Parenti\inst{\ref{i:ias}}
        \and
        O. Podladchikova\inst{\ref{i:pmod}}
        \and
        R. Aznar Cuadrado\inst{\ref{i:mps}}
        \and
        É. Buchlin\inst{\ref{i:ias}}
        \and
        L. Dolla\inst{\ref{i:rob}}
        \and
        C. Verbeeck\inst{\ref{i:rob}}
        \and
        S. Gissot\inst{\ref{i:rob}}
        \and
        L. Teriaca\inst{\ref{i:mps}}
        \and
        M. Haberreiter\inst{\ref{i:pmod}}
        \and
        A.~C.~Katsiyannis\inst{\ref{i:rob}}
        \and
        L. Rodriguez\inst{\ref{i:rob}}
        \and
        E. Kraaikamp\inst{\ref{i:rob}}
        \and
        P.J. Smith\inst{\ref{i:mssl}}
        \and
        K. Stegen\inst{\ref{i:rob}}
        \and
        P. Rochus\inst{\ref{i:csl}}
       \and
        J. P. Halain \inst{\ref{i:csl},\ref{i:esa}}
       \and
        L. Jacques\inst{\ref{i:csl}}
        \and
        W.T. Thompson\inst{\ref{i:gsfc}}              
        \and
        B. Inhester \inst{\ref{i:mps}}          
    }
    \institute{
            Solar-Terrestrial Centre of Excellence -- SIDC, Royal Observatory of Belgium, Ringlaan -3- Av. Circulaire, 1180 Brussels, Belgium\label{i:rob}
            \and
            Université Paris-Saclay, CNRS, Institut d'Astrophysique Spatiale, 91405, Orsay, France\label{i:ias}
            \and
            UCL-Mullard Space Science Laboratory, Holmbury St.\ Mary, Dorking, Surrey, RH5 6NT, UK\label{i:mssl}
            \and
            Max Planck Institute for Solar System Research, Justus-von-Liebig-Weg 3, 37077 G\"ottingen, Germany\label{i:mps}
            \and
            Physikalisch-Meteorologisches Observatorium Davos, World Radiation Center, 7260, Davos Dorf, Switzerland\label{i:pmod}
            \and
            ETH-Z\"urich, Wolfgang-Pauli-Str. 27, 8093 Z\"urich, Switzerland\label{i:eth}
            \and
            Adnet Systems Inc., NASA Goddard Space Flight Center, Code 671, Greenbelt, MD 20771, United States of America\label{i:gsfc}
            \and
            Institute of Applied Computing \& Community Code, Universitat de les Illes Balears, 07122 Palma de Mallorca, Spain\label{i:iaccc}
            \and
            Skobeltsyn Institute of Nuclear Physics, Moscow State University, 119992 Moscow, Russia\label{i:sinp}
            \and
            Department of Mathematics, Physics and Electrical Engineering, Northumbria University, Newcastle Upon Tyne, NE1 8ST, United Kingdom\label{i:north}
            \and
            Centre Spatial de Li\`ege, Universit\'e de Li\`ege, Av. du Pr\'e-Aily B29, 4031 Angleur, Belgium\label{i:csl}
            \and
            European Space Agency (ESA/ESTEC),  Keplerlaan 1, PO Box 299 NL-2200 AG Noordwijk, The Netherlands\label{i:esa}
            \and
            Institute of Geodynamics of the Romanian Academy, Bucharest, Romania\label{i:igar}
    }

   \date{Received January 1X, 2021; accepted February XX, 2021}


  \abstract
   {Heating of the solar corona by small heating events requires an increasing number of them at progressively smaller length scales, with the bulk of the heating occurring at currently unresolved scales.}
   {The goal of this paper is to study the smallest brightening events observed in the extreme UV quiet Sun.}
   {We use commissioning data taken by the EUI instrument onboard the recently launched Solar Orbiter mission. On 2020 May 30, EUI was situated at 0.556\,AU from the Sun. Its \hrieuv telescope (\SI{17.4}{\nano\metre} passband) reached an exceptionally high two-pixel spatial resolution of 400~km. The size and duration of small-scale structures is determined in the \hrieuv data, while their height is estimated from triangulation with the simultaneous SDO/AIA data. This is the first stereoscopy of small scale brightenings at high resolution.}
   {We observed small localised brightenings (``campfires'') in a quiet Sun region
   with lengthscales between \SI{400}{\kilo\metre} and \SI{4000}{\kilo\metre} and durations between  \SI{10}{\second} and \SI{200}{\second}.
   The smallest and weakest of these \hrieuv brightenings have not been observed before. Simultaneous \hrilya observations do not show localised brightening events, but the locations of the \hrieuv events correspond clearly to the chromospheric network.
   Comparison with simultaneous AIA images shows that most events can also be identified in the \SI{17.1}{\nano\metre}, \SI{19.3}{\nano\metre}, \SI{21.1}{\nano\metre}, and \SI{30.4}{\nano\metre} passbands  of AIA, although they appear weaker and blurred.  DEM analysis indicates coronal temperatures peaking at  $\log T\approx6.1 - 6.15$. We determined the height of a few campfires, which is between 1000 and 5000~km above the photosphere. }
{We conclude that ``campfires'' are mostly coronal in nature and are rooted in the magnetic flux concentrations of the chromospheric network. We interpret these events as a new extension to the flare/microflare/nanoflare family. Given their low height, the EUI ``campfires'' could be a new element of the  fine structure of the transition region/low corona:  apexes of small-scale loops that are internally heated to coronal temperatures.}


\maketitle

%

\section{Introduction}

 The Extreme Ultraviolet Imager \citep[EUI, ][]{EUI} onboard Solar Orbiter \citep{SolarOrbiter} consists of three telescopes: the Full Sun Imager (FSI), that operates in two extreme UV bandpasses (\SI{17.4}{\nano\metre}, and \SI{30.4}{\nano\metre}), and two High Resolution Imagers (HRIs).
 The bandpass of `\hrieuv', centered at \SI{17.4}{\nano\metre},  is dominated by \mbox{Fe IX} and \mbox{Fe X} emission at 1~MK. The  bandpass of `\hrilya', centered at \SI{121.6}{\nano\metre},  is dominated by the Lyman-$\alpha$ line of hydrogen.

 The scientific goal of the two HRIs is to observe the fine structure and dynamics, respectively in the upper chromosphere (\hrilya) and in the low corona (\hrieuv).  The transition region between these two layers is where the temperature increases abruptly by two orders of magnitude and where the replenishment of energy and mass to sustain the higher corona, and hence the whole heliosphere, takes place. Its physics is very complex, and it is clear that the crucial processes responsible for the formation of the corona must be operating at all times at very small spatial and temporal scales \citep{Parker1988, Klimchuk2015}.

\begin{figure*}[ht]
\centering
\includegraphics[width=1.0\hsize]{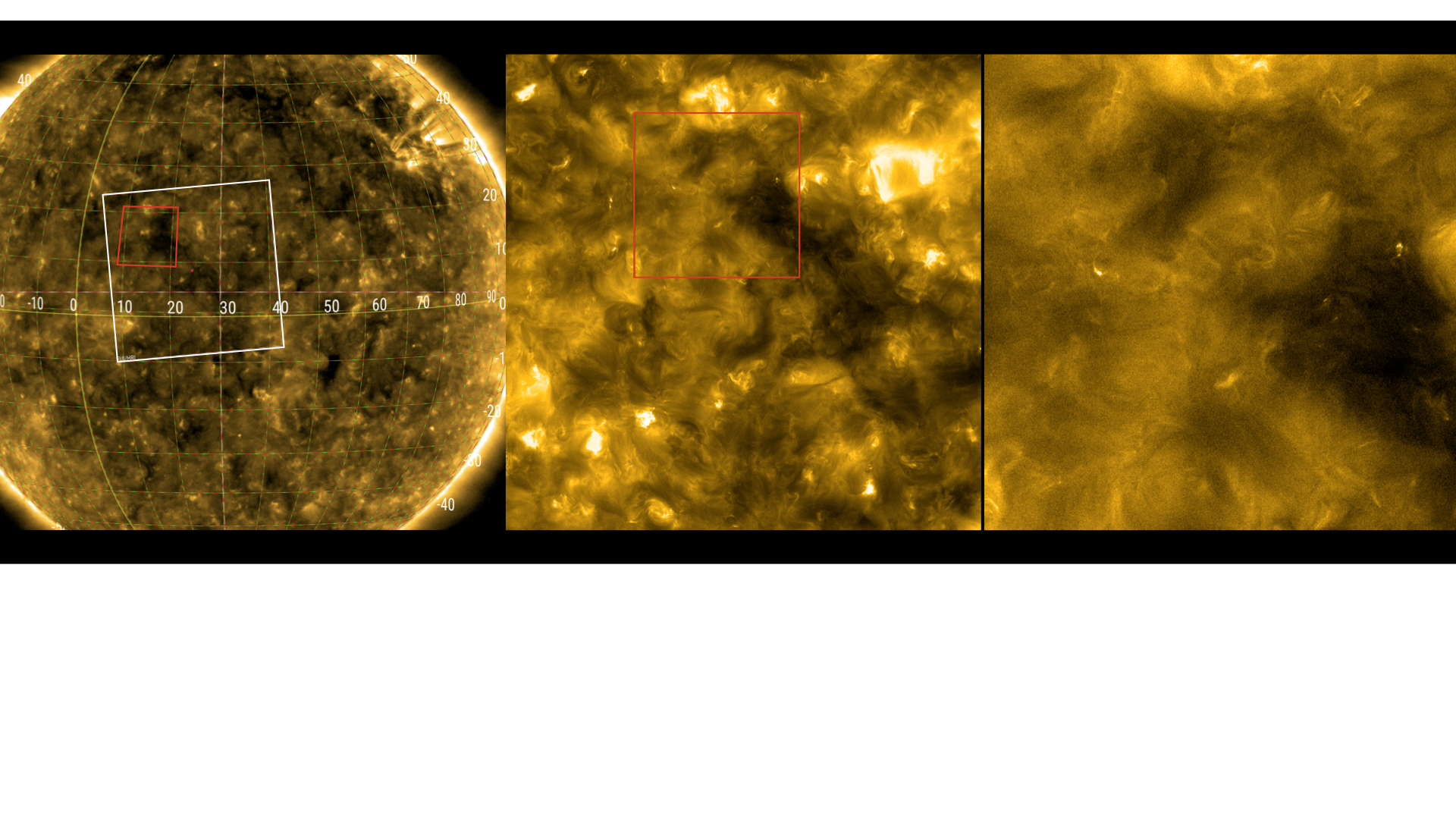}
\caption{(Left) The location of the HRI field of view is shown as a white square on an EUI/FSI \SI{17.4}{\nano\metre} image. Stonyhurst coordinates are shown, with the crossing of the two yellow great circles indicating the Earth/SDO subsolar point.
sNote that the spacecraft roll angle is aligned with the North direction of the Solar Orbiter orbital plane, not with the solar meridional plane.
Image visualisation by JHelioviewer \citep{Mueller2017}. (Middle) The first image of the \hrieuv sequence. (Right) subfield corresponding to the red square indicated in the left two panels. The brighthest features correspond to the "campfires" studied in this paper.}
\label{fig:scene}
\end{figure*}

The classical picture of the transition region is that of a thermal interface between the coronal and chromospheric plasma \citep{Gabriel1976}. It, however, cannot describe properly high radiance of the transition region lines, so the concept of the ``unresolved fine structure'' (UFS) of the transition region in the quiet Sun has been proposed \citep[see][]{Feldman1983, Dowdy1986}: a plethora of small-scale low-lying loops that are magnetically isolated from the corona. Such structures were later detected with the improvement of the spatial resolution of observations \citep[e.g.][]{Feldman1999, Warren2000}. Recently, \cite{Hansteen2014} used the data from the IRIS mission (\cite{IRIS}, spatial resolution of
\SI{0.4}{\arcsecond}, \SI{290}{\kilo\metre})
and found that low-lying loops
(projected full lengths of 4-12~Mm) at low transition-region temperatures ($\log T $ around 4.3--4.8) to be a major constituent of the UFS. The Hi-C sounding rocket \citep{HIC,Rachmeler2019} reached the highest resolution in extreme UV coronal observations so far (\SI{0.3}{\arcsecond}, \SI{218}{\kilo\metre}) and detected fine scale features of less than \SI{200}{\kilo\metre} wide \citep{Peter2013, Barczynski2017}. However, up to now Hi-C only observed active regions, so the fine structure of the upper transition region of the quiet Sun remains unexplored.

Earlier studies of the quiet Sun have  revealed a wide range of extreme UV  brightenings at transition region and low coronal temperatures, often interpreted in terms of nanoflare heating. We use the term nanoflare here to refer to the energy range, regardless of the driving mechanism. For coronal heating purposes, an increasing number of miniature heating events are required at progressively smaller length scales, with a power law distribution with an index steeper than $-2$ \citep{Hudson1991, Berghmans2002, Joulin2016}.  A power-law distribution has been found for the number of events as a function of their energy content \citep{Benz2002, Aschwanden2004}. A power-law index close to $-2$ has been found  by \cite{Berghmans1998}, whereas a power-law index of $-2.5$ was found by \cite{Harra2000}.
However, \cite{Aschwanden2002} found a power law of index $-1.54$, which matches the distribution of active region transient brightenings \citep{Shimizu1994} and hard X-ray flares \citep{Crosby1993},  and explained the difference in terms of a temperature bias imposed by narrow band extreme UV imagers.
Recently, \cite{Chitta2021} analysed `EUV bursts' and concluded that, at the resolution of  SDO/AIA,   there were by a factor 100 insufficient events to account for the energy that is required to heat the corona.
At transition region temperatures, explosive events are seen in spectroscopic data with lifetimes of order of minutes and sizes below  \SI{1}{\mega\metre} \citep{Brueckner1983, Dere1989, Teriaca2004}.

This work uses the first release of properly calibrated and formatted EUI data in December 2020. Before that date, only uncalibrated, quicklook EUI data were publicly available for outreach events such as the ESA press release in July 2020 were the EUI team first presented campfires. The objective of this Letter is to present the novel observations of the EUI/HRIs (Section \ref{section:Observations}),  and discuss what these can contribute to our understanding of coronal structuring and heating (Section \ref{section:Discussion}).

\section{Observations}
\label{section:Observations}

As part of the instrument commissioning, the two EUI HRI telescopes were operated on 2020 May 30 for the first time in parallel at a fast imaging cadence of 5s,  and acquired 50 images each.  Solar Orbiter was located at \ang{31.5} west in solar longitude from the Earth-Sun line (see Fig.~\ref{fig:scene}). In this study we will also use the data acquired by the AIA instrument \citep{AIA} onboard the Solar Dynamics Observatory \citep[SDO, ][]{SDO} in Earth orbit.  To compare EUI and AIA across the whole FOV, we have remapped the images from both instruments (see Appendix \ref{section:dataset}) to a  regular $2400\times2400$ grid in Carrington coordinates with a 0.01625 heliographic degrees (197 km) pitch, which corresponds to the original \hrieuv resolution. Appendix \ref{section:dataset} contains the details of the acquired dataset.

In Fig.~\ref{fig:scene} we show the \hrieuv field of view. The subject of the present study are the small-scale brightenings appearing in \hrieuv images with sizes down to the 2-pixel resolution limit of the telescope. These brightenings have become known as ``campfires'', a term which we  use for ``small  brightenings observed by \hrieuv'' without necessarily implying that these are physically different from previously observed small extreme UV brightenings in the corona.

\begin{figure*}[ht]
\centering
{ \includegraphics[width=1.0\hsize]{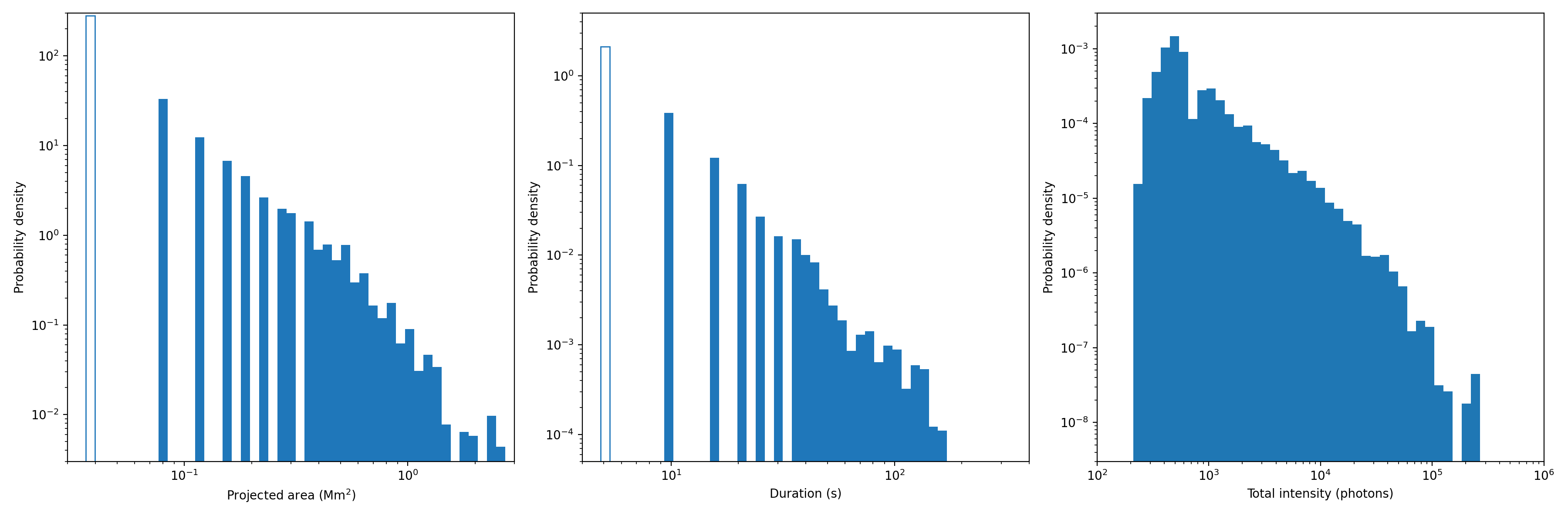}}
\caption{The observed event distribution as a function of projected area, duration and total intensity. Projected area refers to all the pixel area that the events occupy at some stage over their duration. The probability density is scaled such that when integrated over all bin widths, the total event count of 1467 is obtained. The smallest areas and durations (only one pixel or one exposure) are displayed as bars without filling.
}
\label{fig:histograms}
\end{figure*}

We used an automated detection method to determine the statistical significance of small features spatially and track them in time through the sequence. The detection scheme (see Appendix \ref{Automated detection}) applied to the stack of 50 Carrington projected \hrieuv images revealed 1467 campfires. Dividing this number by the area of the field of view and by the duration of our sequence, we arrive at a birthrate of
$3.7\times 10^{-17} \mbox{m}^{-2} \mbox{s}^{-1}$, which is an order of magnitude lower than the birthrate of explosive events (\cite{Teriaca2004} found $4.1\times 10^{-16} \mbox{m}^{-2} \mbox{s}^{-1}$).

Fig.~\ref{fig:histograms} shows the histograms of the projected area, duration, and total intensity of these events.

Note that we did not remove the left-most bins of the histograms in Fig.~\ref{fig:histograms} that correspond to events that appear only in 1 image or only in 1 pixel. Since these bins are in agreement with the trend seen at larger scales, we kept them in the figure
as an illustration that the campfires population extends consistently down to the limitations of our data set in terms of spatial and temporal resolution. Individual events of 1 pixel in 1 frame cannot be confirmed but some 1-pixel events can last several frames and some 1-frame events consist of several pixels.
Out of the 1467 events, there are 558 which have both their projected area larger than 1 pixel and their duration larger than 5 s (the imaging cadence).

The detected population is limited by the total duration of our \hrieuv  data set, which is only 245~s. The longest event durations in our data set approaches 200~s, but among these, many events are truncated at the end or the start of the sequence. There is no such artificial truncation in the histogram of the projected area, where the largest events are on the order of a few Mm$^2$, which corresponds to roughly one hundred pixels. Strikingly, both the histogram of projected area and the histogram of campfire duration show a powerlaw-like behaviour
stretching over at least 2 orders of magnitude.

Many events have an elongated nature, reminiscent of small loop-like structure. Linear (projected) lengths of these elongated structures range from 1 to 20 pixels (\SI{0.2}{\mega\metre} to \SI{4}{\mega\metre}) with aspect ratios  distributed mostly between 1 and 5. The smallest observed structures are similar to the smallest active region loops observed by Hi-C \citep{Peter2013}.

\begin{figure*}[h]
\centering
{ \includegraphics[width=1.0\hsize]{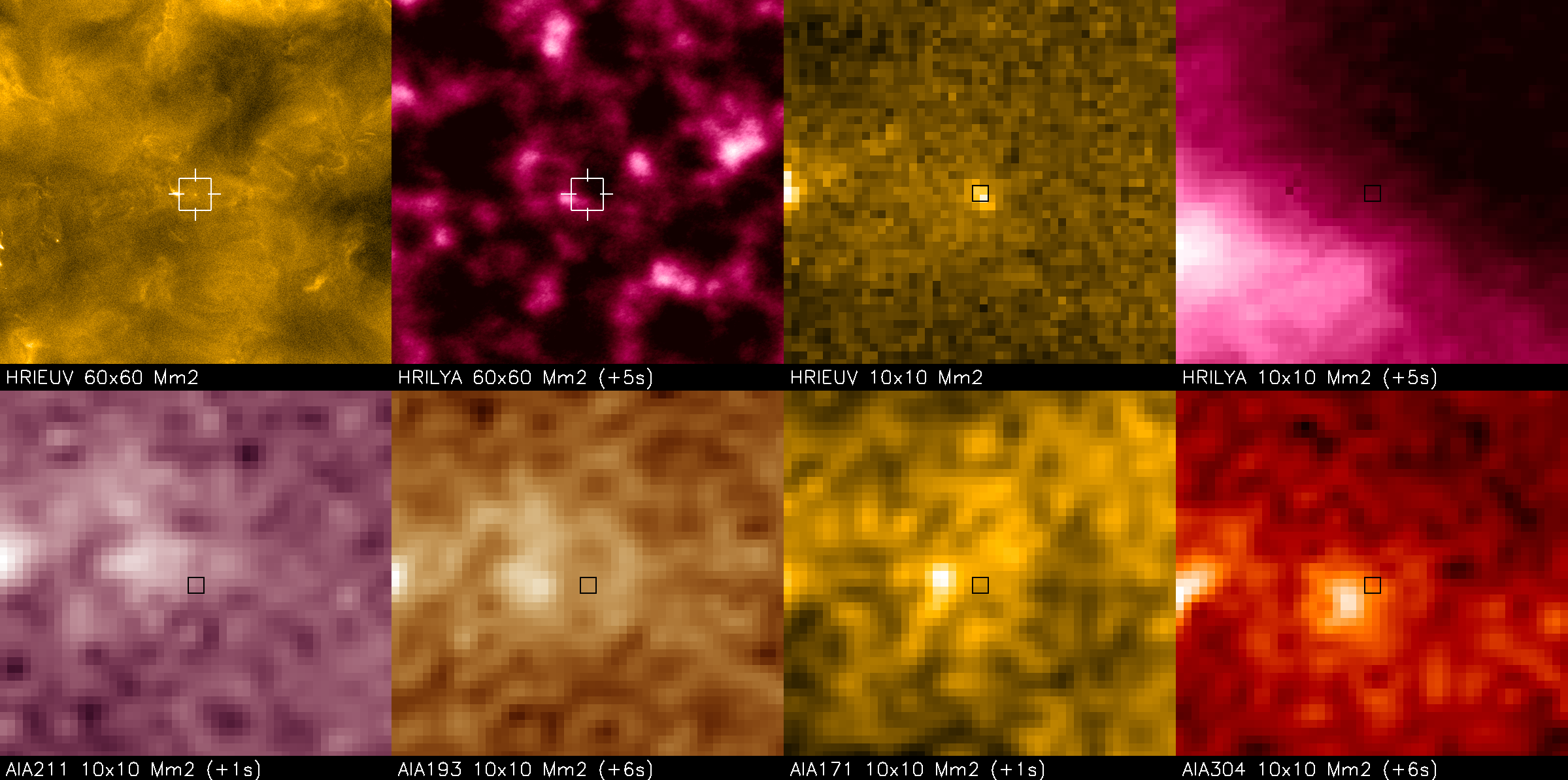}} 
\caption{Example of a dot-like event. The white square in the two top-left panels corresponds to the field of view of the other, zoomed-in panels.
}
\label{fig:cf344_BP6}
\end{figure*}

\begin{figure*}[h]
\centering
{ \includegraphics[width=1.0\hsize]{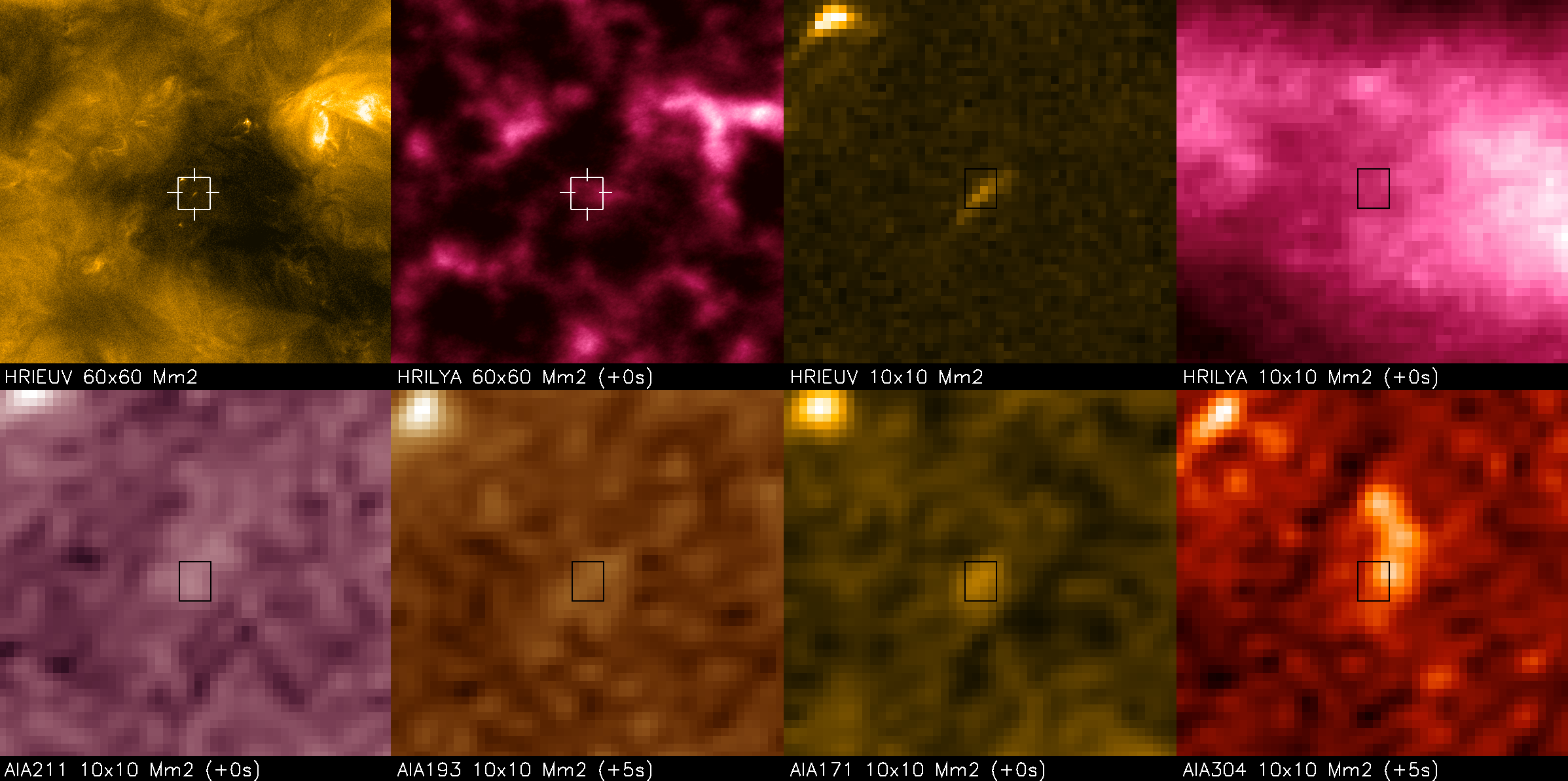}} 
\caption{Example of a loop like event. The white square in two the top-left panels corresponds to the field of view of the other, zoomed-in panels.
}
\label{fig:cf297_BP12}
\end{figure*}

As expected from the histograms in Fig.~\ref{fig:histograms}, most campfires are very close to the \hrieuv resolution limit and are therefore either dot-like (Fig.~\ref{fig:cf344_BP6}) or short loop-like (Fig.~\ref{fig:cf297_BP12}).  In most cases, these small events can be recognised in the SDO/AIA bandpasses  at \SI{21.1}{\nano\metre}, \SI{19.3}{\nano\metre}, \SI{17.1}{\nano\metre}, and \SI{30.4}{\nano\metre}, although they typically appear weaker and fuzzier as can be expected from the difference in spatial resolution. No clear event signatures were identified in any of the other SDO/AIA bandpasses. In Appendix~\ref{section:Multiwavelength} we show a few more examples, including bigger events with internal structure such as possibly interacting loops.

\begin{figure}[h]
\centering
{ \includegraphics[width=0.9\hsize]{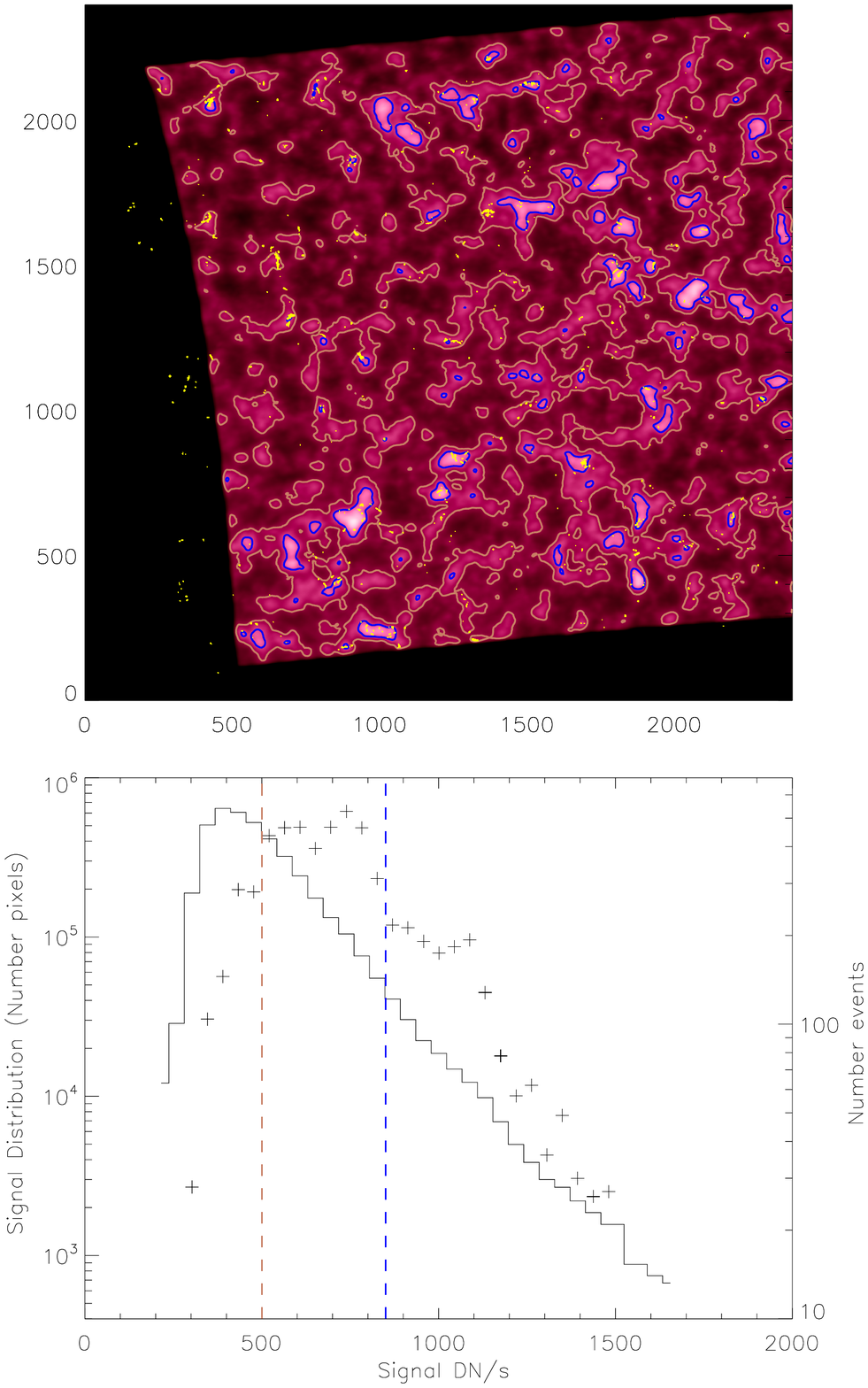}}
\caption{
Top: Averaged intensity map of 50  \hrilya images. Detected ``campfire'' events are displayed in yellow. Isocontours correspond to signals of 500 DN/s (brown) and 850 DN/s (blue).
Bottom: Intensity distribution of the  \hrilya averaged map (solid histogram). The number of detected ``campfire'' events present in each intensity bin is represented with crosses. Vertical dashed lines embed signals between 500 DN/s (brown) and 850 DN/s (blue), where the majority of the detected events are present. Both y-axes are displayed in logarithmic scale.}
\label{fig:hrilya_distribution}
\end{figure}

Remarkably, the observed campfires do not reveal unambiguously localised brightenings in \hrilya at the same location and time as the campfires detected in \hrieuv. Possibly the \hrilya is of a more fuzzy, distributed brightening of a bigger neighborhood, which will be studied in future work. In any case, the  \hrieuv campfires have a clear tendency to outline the chromospheric network as shown by \hrilya. In the top panel of Figure~\ref{fig:hrilya_distribution}, we show the location of the  campfires detected in the \hrieuv data,  over-plotted as yellow dots  on a reference \hrilya image, which was obtained by averaging over the 50 Carrington projected  \hrilya images.


The intensity distribution of the averaged  \hrilya map is shown in the bottom panel of Figure~\ref{fig:hrilya_distribution} (solid histogram). The number of detection events falling within each of the 50 equal intensity bins is displayed with crosses.
 The vertical dashed lines are arbitrarily chosen intensity levels encompassing the peak of the event distribution (crosses) and are used as isocontours on the intensity map (top panel). The brown and blue colours correspond in both panels to the same intensity levels. The detected events are generally not in the darker cell-centers and there are also very few of them at the brightest locations. The majority of the campfires are  located at the network boundaries, which is the same location as where explosive events are found \citep{Porter1991}. Note that, in contrast to the SDO/AIA images, the \hrilya images are taken from exactly the same perspective as \hrieuv. Projection effects are therefore not likely to have a major influence on the apparent location of \hrieuv events in the \hrilya images.

\begin{figure}[h]
\centering
{ \includegraphics[width=0.8\hsize]{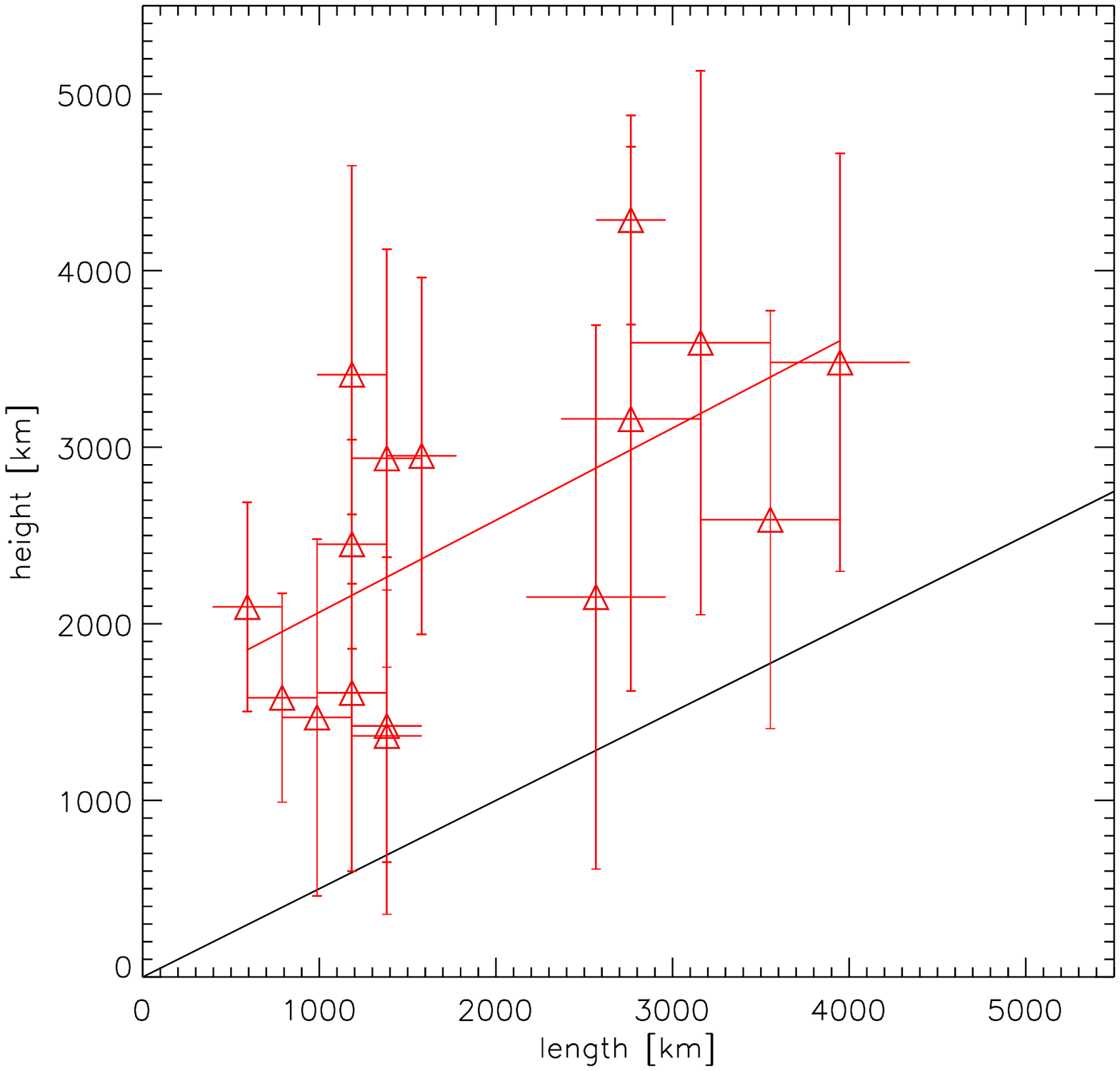}}
\caption{
The height above the photosphere of the 16 campfires versus their projected length (red triangles with their error bars, $h \pm \delta h$).
The red line is a linear fit to the data points.
The black line represents the height as half of the projected length as one would expect for a semi-circular loop, for comparison.}
\label{fig:height_size}
\end{figure}

In order to determine the height of campfires above the photospheric surface, we performed triangulation on sixteen small campfires that could be identified on both \hrieuv  \SI{17.4}{\nano\metre} and AIA \SI{17.1}{\nano\metre} images.  Figure~\ref{fig:height_size} shows the height above the photosphere versus the projected length for 16 campfires selected for an easy identification. The projected length of each campfire is calculated as the distance between the farthest two pixels of the campfire in the image plane. The photospheric radius value of 695700~km \citep{Prsa2016} is subtracted to obtain the height above the photosphere. The error bars are derived in  Appendix ~\ref{section:Height of campfires}. From Figure~\ref{fig:height_size} one can see that campfires are located between 1000~km and 5000~km above the photospheric surface.

\begin{figure}[h]
\centering
{ \includegraphics[width=1.0\hsize]{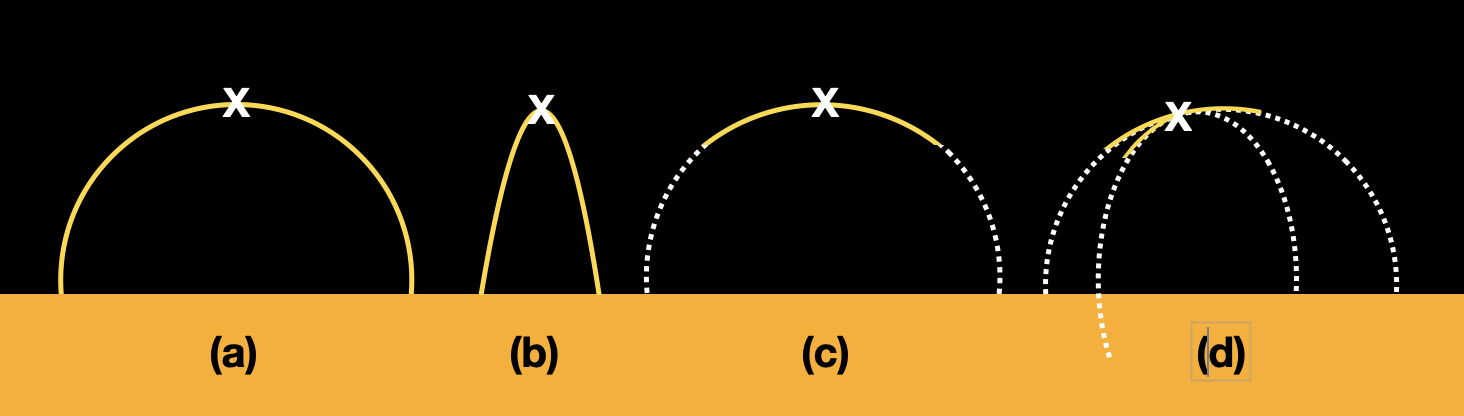}}
\caption{
\textbf{Scenarios to interpret the campfire heights shown in Fig. \ref{fig:height_size}. The yellow solid lines are the observed campfires, the dotted lines are the invisible loop segments, and the white cross is the triangulation point. }
}
\label{fig:loopheight_cartoon}
\end{figure}

The 3D geometry of campfires is unclear as many of them are barely resolved.  Various complex geometries could be envisaged, but here we start with the simplest assumption that campfires are semi-circular loops.
A semi-cicular loop has an  apex at height $H$ which corresponds to half the length between the loop footpoints $L$.  The black $H  = L/2$ line in Figure~\ref{fig:height_size} is, per definition of apex, the upper limit for the height of any point on a semi-circular loop. For our 16 campfires we triangulated the brightest pixel, or the mid-point of a few bright pixels if there are several adjacent pixels of comparable brightness.
Figure~\ref{fig:height_size} shows that the resulting  campfire heights are systematically higher than the $H  = L/2$ line.  Figure~\ref{fig:loopheight_cartoon} shows possible interpretations.
A semi-circular loop (a) clearly does not explain the height observations.
If campfires represent small-scale loops, then they are either (b) elongated (tall) loops, or only their apexes are visible (c). The latter possibility seems more likely as we do not expect the \SI{17.4}{\nano\metre} emission to extend all the way down to the photosphere. Such a scenario is conceivable when e.g. the observed campfire corresponds to emission near a contact point of nearby loops (d).

\begin{figure}[h]
\centering
{ \includegraphics[width=1.0\hsize]{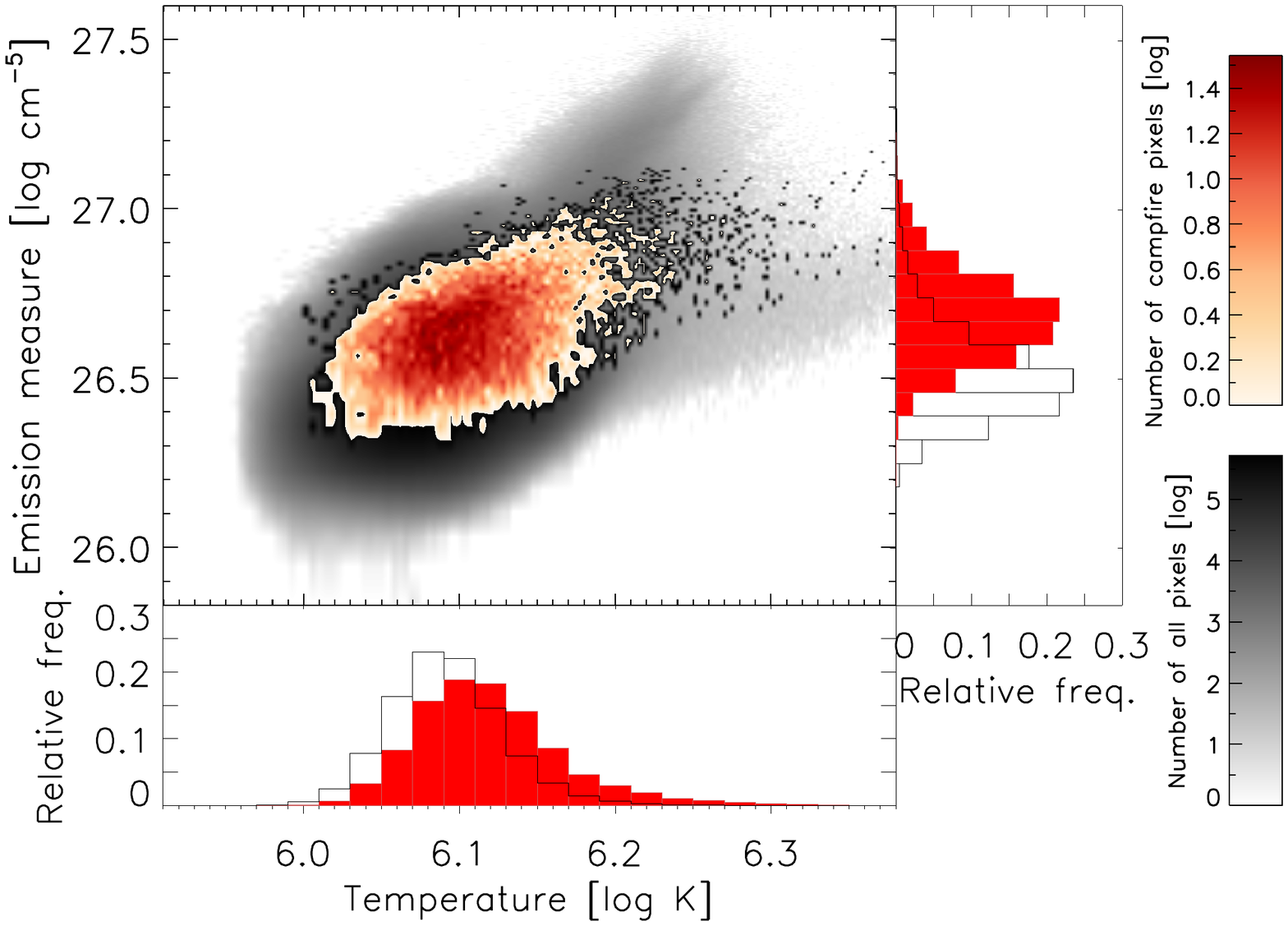}}
\caption{2D histogram of the total emission measure and DEM-weighted temperature for campfire pixels (red) and all pixels (black). The color is given according to the logarithm of the number of pixels. The number of campfire pixels and all pixels are roughly 9800 and $1.1\times10^8$, respectively.
}
\label{fig:EM_versus_temp}
\end{figure}

The temperatures of the events were estimated through Differential Emission Measure (DEM) analysis, using the methods of \citet{Hannah_2012AA...539A.146H} and \citet{Cheung_2015ApJ...807..143C}. Here we show only the result obtained by the former method. The temperatures inferred this way are essentially the same as for the other method. In Fig.~\ref{fig:EM_versus_temp} we plot the 2D histogram of the total emission measure (EM) for each pixel (integrated over temperature) and the DEM-weighted temperature, as defined in \citet{Parenti_2017}. We exclude campfires with only 1 pixel or lasting only 1 frame. We also correct for the shift in location caused by the different viewpoints by interpolating each event to the location of its intensity barycentre. The temperature distribution peaks at $\log T\approx6.1$, with a FWHM of $5.43$ ($\log$~K), and with a relatively long tail extending to 2~MK.  While this strongly indicates that campfires are coronal events,  i.e. that the plasma in these brightenings reaches temperatures of 1~MK and more, it should be noted, however, that campfires may not always dominate the EM along the LOS, and therefore, that the DEM-weighted temperature may not always be a good representative of the temperatures. Nonetheless, we note that both the emission measure and the temperature distribution of the events, but in particular the former, are clearly shifted towards higher values with respect to the background.

\section{Summary and Discussion}
\label{section:Discussion}

In a first high-cadence dataset from the high-resolution telescopes of EUI, we have identified a large number of extreme UV brightenings in the quiet Sun that have been called ``campfires''.  To our knowledge, this is the highest resolution image sequence of the quiet Sun corona ever. Given the favorable \SI{31.5}{\degree} separation in heliocentric longitude with SDO/AIA in Earth orbit, we could also apply triangulation techniques at the highest resolution ever to determine the height of campfires. The characteristics of campfires are summarised below.

 An increasing number of campfires is found at progressively smaller scales, down to the resolution of the present \hrieuv dataset. We find apparent powerlaw distributions for the campfire areas, durations and total intensity.  The area distribution is limited at the lower end by spatial resolution (a 2-pixel area is  $ \SI{0.08}{\mega\metre}^2$). At the higher end, the area distribution ends around $\SI{5}{\mega\metre}^2$, which is a consequence of our wavelet based detection scheme. The duration distribution is limited at the lower end by the temporal resolution of the present dataset (twice the image cadence is \SI{10}{\second}). At the higher end, the duration distribution ends around $\SI{200}{\second}$, which  is close to the total duration of our dataset, namely $\SI{245}{\second}$.
 The appearance of the campfires at progressively smaller scales ranges from small systems of perhaps interacting loops, to individual loops and finally dots at the resolution limit of the data set. The  aspect ratios of length versus width  range from 1 to  5.

 The observed campfires observed in \hrieuv  do not have obviously corresponding localised brightenings in \hrilya at the same time and place. However, the events detected in \hrieuv do line up along the chromospheric network as observed by  \hrilya. Comparison with simultaneous AIA images shows that most events can be found back in \SI{21.1}{\nano\metre}, \SI{19.3}{\nano\metre}, \SI{17.1}{\nano\metre} and \SI{30.4}{\nano\metre}  too, although as compared to \hrieuv, they appear weaker and fuzzier as expected from the difference in resolution.  They are not seen in `hotter' AIA bandpasses.

Using triangulation techniques it was determined for a sub-sample of events that the campfires are located between \SI{1000}{\kilo\metre} and \SI{5000}{\kilo\metre} above the photosphere.  Campfires are events at coronal temperatures: the distribution of their DEM-weighted temperatures is peaked at $\log T\approx6.12$, with a FWHM of $0.074$ ($\log$~K).

Using Hi-C data, \citet{Tiwari2019} reported energy-release events that resemble some of the bigger campfires observed by \hrieuv, but they only analyzed active region observations.
In the quiet Sun, the smallest  extreme UV brightenings have been observed with SDO/AIA \citep{Joulin2016, Chitta2021}. The superior spatial and temporal resolution of the studied \hrieuv quiet Sun data set allows us to extend the family of extreme UV quiet Sun brightenings down to spatial scales less than half a Mm, and down to temporal scales of 10~s.

Campfires are located above the chromospheric network, where the strongest quiet Sun magnetic fields are. This suggests that campfires are magnetically defined features, as is further supported by their often isolated, distinct loop-like appearance.
While the campfires do reach coronal temperatures  (1--1.6~MK), they are located  at remarkably low heights, between 1000~km and 5000~km from the  photosphere.
We believe it is the first time such low heights of quiet sun structures emitting around \SI{17.4}{\nano\metre} have been determined.
Compared to their lengths, the campfires are systematically higher than what could be expected for semi-circular loops. Either campfires are vertically elongated loops, or campfires correspond only to loop tops brightening up.  We propose that campfires are the apexes of  low-lying small-scale network loops that are heated to coronal temperatures.

The observed heights of campfires at coronal temperatures coincide with the heights of the cold transition region loops reported by \citet{Hansteen2014}.
This co-existence is coherent with the picture of a plethora of small-scale low lying loops existing at different temperatures  \citep{Feldman1983, Dowdy1986}. Future research will address the question for what reason the campfire loops stand out as particularly hot when compared to neighbouring loops that do not surpass transition region temperatures.

The bigger campfires have durations and sizes comparable to explosive events \citep{Dere1989}, even if the latter ones are rather observed at transition region temperatures, see also \citet{Peter2014}, and they both appear at network boundaries. Campfires are  possibly an order of magnitude less frequent than explosive events, though this might be a detection bias as explosive events are spectroscopically identified. Some of the AIA \SI{17.1}{\nano\metre} quiet Sun small brightenings studied by \cite{Innes2013} are clearly associated with spectral profiles typical of explosive events. Due to the lack of spectroscopic data, this could unfortunately not be checked for the campfire population.

The current observations where taken at a distance of 0.556~AU from the Sun. Future perihelia of Solar Orbiter will go down to below 0.3~AU and increase the effective spatial resolution of \hrieuv with another factor two.
Joint campaigns with other remote sensing instruments onboard Solar Orbiter such as the Polarimetric and Helioseismic Imager \cite[PHI,][]{PHI} and the SPICE spectrograph \citep{SPICE} will help confirm or falsify our interpretation. This is particularly important for determining the driver of these heating events, their link to explosive events and their role in coronal heating.


\begin{acknowledgements}
  The building of EUI was the work of more than 150 people during more than 10 years, we gratefully acknowledge all the efforts that have led to a successfully operating instrument.
  The authors thank the Belgian Federal Science Policy Office (BELSPO) for the provision of financial support in the framework of the PRODEX Programme of the European Space Agency (ESA) under contract number 4000112292. The French contribution to the EUI instrument was funded by the French Centre National d'Etudes Spatiales (CNES); the UK Space Agency (UKSA); the Deutsche Zentrum f\"ur Luft- und Raumfahrt e.V. (DLR); and  the Swiss Space Office (SSO). PA and DML acknowledge funding from STFC Ernest Rutherford Fellowships No. ST/R004285/2 and ST/R003246/1, respectively. SP acknowledges the funding by CNES through the MEDOC data and operations center.
\end{acknowledgements}

\bibliographystyle{aa} 
\bibliography{EUI_bibliography,SO_Book_cross_references} 
\begin{appendix}

\section{Dataset description}
\label{section:dataset}

As part of the instrument commissioning, the two EUI HRI telescopes were operated for the first time in parallel at a fast imaging cadence on 2020 May 30. Solar Orbiter was located at \ang{31.5} west in solar longitude from the Earth-Sun line (see Fig.~\ref{fig:scene}), and at a distance of 0.556\,AU from the Sun. From this vantage point, the \hrieuv angular pixel size of \SI{0.492}{\arcsecond} corresponds to \SI{198}{\kilo\metre} on the solar surface. The \hrilya angular pixel size of \SI{0.514}{\arcsecond}  corresponds to \SI{207}{\kilo\metre}.
 \hrieuv effectively achieves a two-pixel resolution of \SI{0.984}{\arcsecond}, while for \hrilya, diffraction from the 30~mm aperture and the type of detector limited the goal to \SI{2}{\arcsecond}, with an effective resolution estimated now of about \SI{3}{\arcsecond}. Both HRI cameras produce $2048\times2048$ pixel images that correspond to a \SI{16.8 x 16.8}{\arcmin} FOV.

The data analysed for this study were acquired between 14:54:00\,UT and 14:58:20\,UT and consist of two sequences of 50 images each, taken simultaneously by the two HRIs at \SI{5}{\second} cadence. The sequence of \hrieuv images started \SI{15}{\second} earlier. \hrieuv images were acquired with \SI{3}{\second} exposure while the  \hrilya images were exposed for \SI{1}{\second}. The data were acquired with lossless compression and processed to Level 2 FITS files as distributed in the EUI Data Release 1 (\footnote{DOI will be inserted later}).

We have also used the data acquired by the AIA instrument \citep{AIA} onboard the Solar Dynamics Observatory \citep[SDO, ][]{SDO} in Earth orbit. The AIA bandpass at \SI{17.1}{\nano\metre} is similar but not identical to the \hrieuv and FSI bandpasses at \SI{17.4}{\nano\metre}. The detailed differences in bandpasses are not considered here but will be the subject of future investigations (Gissot, Auchère  priv. comm.). The  angular pixel sizes of AIA and \hrieuv are comparable but given the difference in heliocentric distance, EUI provides a factor 2 improvement in spatial resolution. The difference in heliocentric distance results also in difference in light travel time of \SI{220}{\second}, meaning that EUI images acquired at time $T$ are to be compared with AIA images acquired at $T$+\SI{220}{\second}. The AIA imaging cadence is \SI{12}{\second}, hence the closest in time AIA and EUI images can be up to \SI{5}{\second} seconds off. We used despiked AIA images. In order to subtract residual thermal signal in faint regions, we subtracted for each quadrant the mean of a 100$\times$100 pixels box in the corresponding corner.

To compare EUI and AIA across the whole FOV, we will use images from both instruments that are remapped to Carrington coordinates. In this process, the HRI images were re-sampled with bi-cubic interpolation on a regular $2400\times2400$ grid with a 0.0163 heliographic degrees (198 km) pitch, thus preserving the \hrieuv resolution.

No extra data points are created in the conversion from the original $2048\times2048$ to the $2400\times2400$ frame as the extra pixels correspond to a dark border around the actual FOV
(see top frame of Fig. \ref{fig:hrilya_distribution}).  Following an optimisation process, the radius of the Carrington projection sphere was chosen at $1.004\,R_{\rm Sun}, $ or \SI{2.8}{\mega\metre} above the photosphere,  in order to take out the average parallax over the field of view between the campfires seen in \hrieuv and AIA.

Individual events can still display a parallax between their \hrieuv and AIA images when their height deviates from this average. This can be an explanation for the mismatch seen in Fig.~\ref{fig:cf161_BP16} if this campfire is located exceptionally high above the photosphere.

\section{Automated detection}
\label{Automated detection}

We use an automated detection method to separate these space-time events from the random intensity fluctuations caused by the various sources of noise in the acquisition chain. In order to determine the statistical significance of a space-time blob of intensity, one must assume a model of the noise present in the data. In the following, we considered only the photon shot noise, which is expected to dominate the other sources of noise in \hrieuv  \citep{EUI}.  While a detection in space-time directly would be desirable, in the present work we determined the statistical significance of small features spatially and tracked them in time through the sequence. The detection scheme applied to the stack of 50 Carrington projected images was as follows:

\begin{enumerate}
\item
Small features are separated from the photon shot noise using the first two scales of a dyadic ``\`a trous'' wavelet transform of the images using a $B_3$ spline scaling function \citep[e.g., ][]{Starck2002}. In \citep{Starck1994}, the wavelet coefficients at each scale are considered significant when they are greater than $n$ times the rms amplitude expected from Gaussian white noise. In this study, we use the extension to Poisson statistics proposed by \cite{Murtagh1995}, with $n=5$ sigma thresholding. The thresholding of the wavelet coefficients results in a binary cube.

\item
The 6-connected voxels of the binary cube are clustered into numbered regions. Each region defines an event.

\item
For each event, physically relevant parameters are computed, such as projected area, duration, parameters of the ellipse of same second moments.

\end{enumerate}

The above procedure with 2 wavelet scales and $n=5$ sigma thresholding, applied on the sequence of 50 \hrieuv images resulted in 1467 campfires.

\section{Example campfires}
\label{appendix:Example campfires}
\label{section:Multiwavelength}

We compare the morphology of a few individual campfires as observed in \hrieuv, \hrilya and in the SDO/AIA bandpasses  at \SI{21.1}{\nano\metre}, \SI{19.3}{\nano\metre}, \SI{17.1}{\nano\metre}, and \SI{30.4}{\nano\metre}. No clear event signatures were identified in any of the other SDO/AIA bandpasses.

The following figures show examples of individual campfires. In each figure, the top row corresponds to \hrieuv and \hrilya subfield images, while the panels on the second row show SDO/AIA subfield images, from left to right at respectively \SI{21.1}{\nano\metre}, \SI{19.3}{\nano\metre}, \SI{17.1}{\nano\metre} and \SI{30.4}{\nano\metre}. Each panel corresponds to nearly $(\SI{10}{\mega\metre})^2$ on the Sun (50$\times$50 pixels, each with a footprint on the Sun of $(\SI{198}{\kilo\metre})^2$, except the two panels top left in each figure that show the $(\SI{60}{\mega\metre})^2$ neighbourhood in which the campfire takes place.
 The black rectangle shows the maximal extension in the $x$ and $y$ direction over which each campfire is detected over its duration.

\begin{figure*}[h]
\centering
{ \includegraphics[width=1.0\hsize]{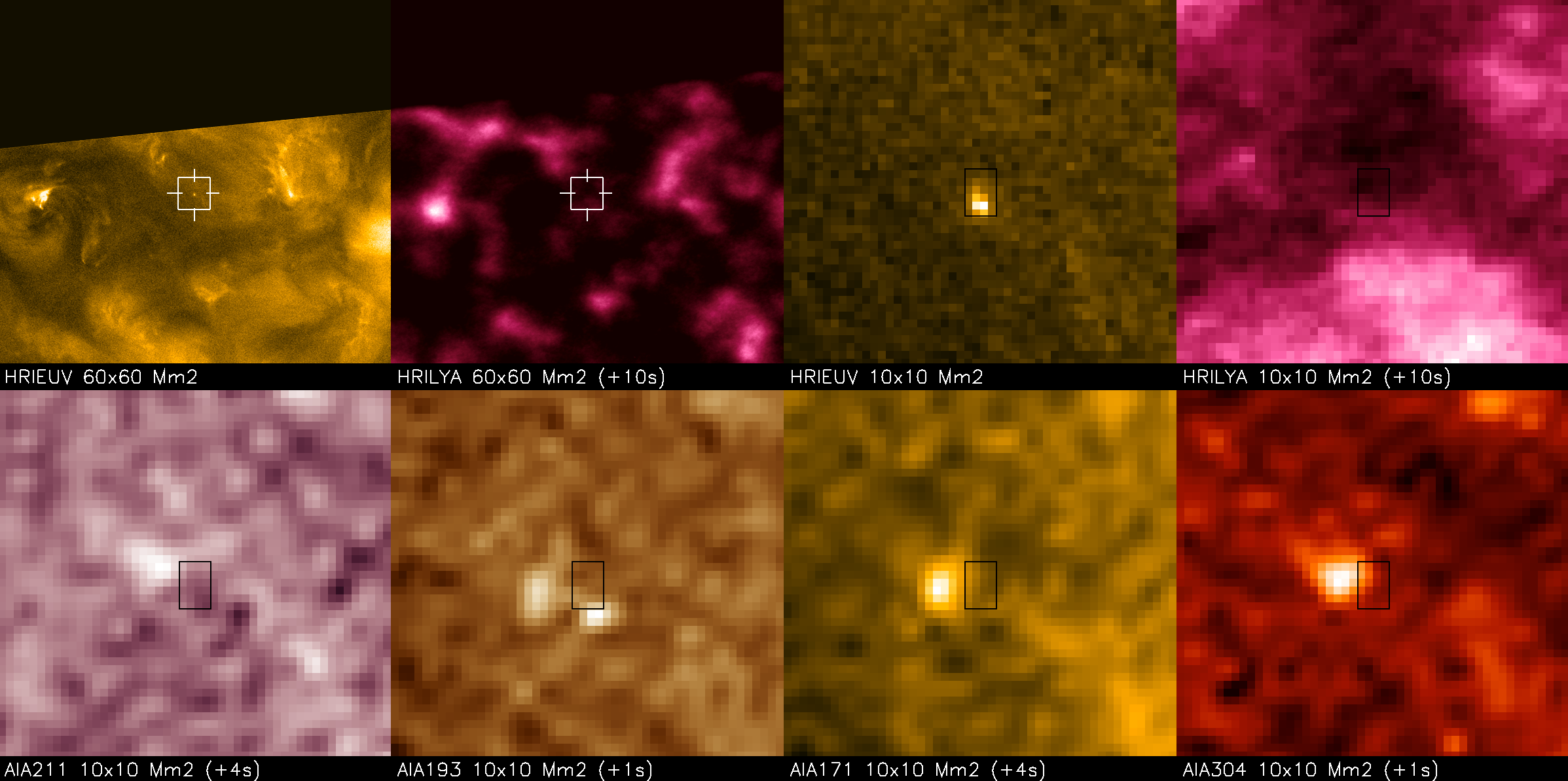}}  
\caption{Example of a dot-like event. The white square in the two top-left panels corresponds to the field of view of the other, zoomed-in panels.
}
\label{fig:cf709_BP2}
\end{figure*}

\begin{figure*}[h]
\centering
{ \includegraphics[width=1.0\hsize]{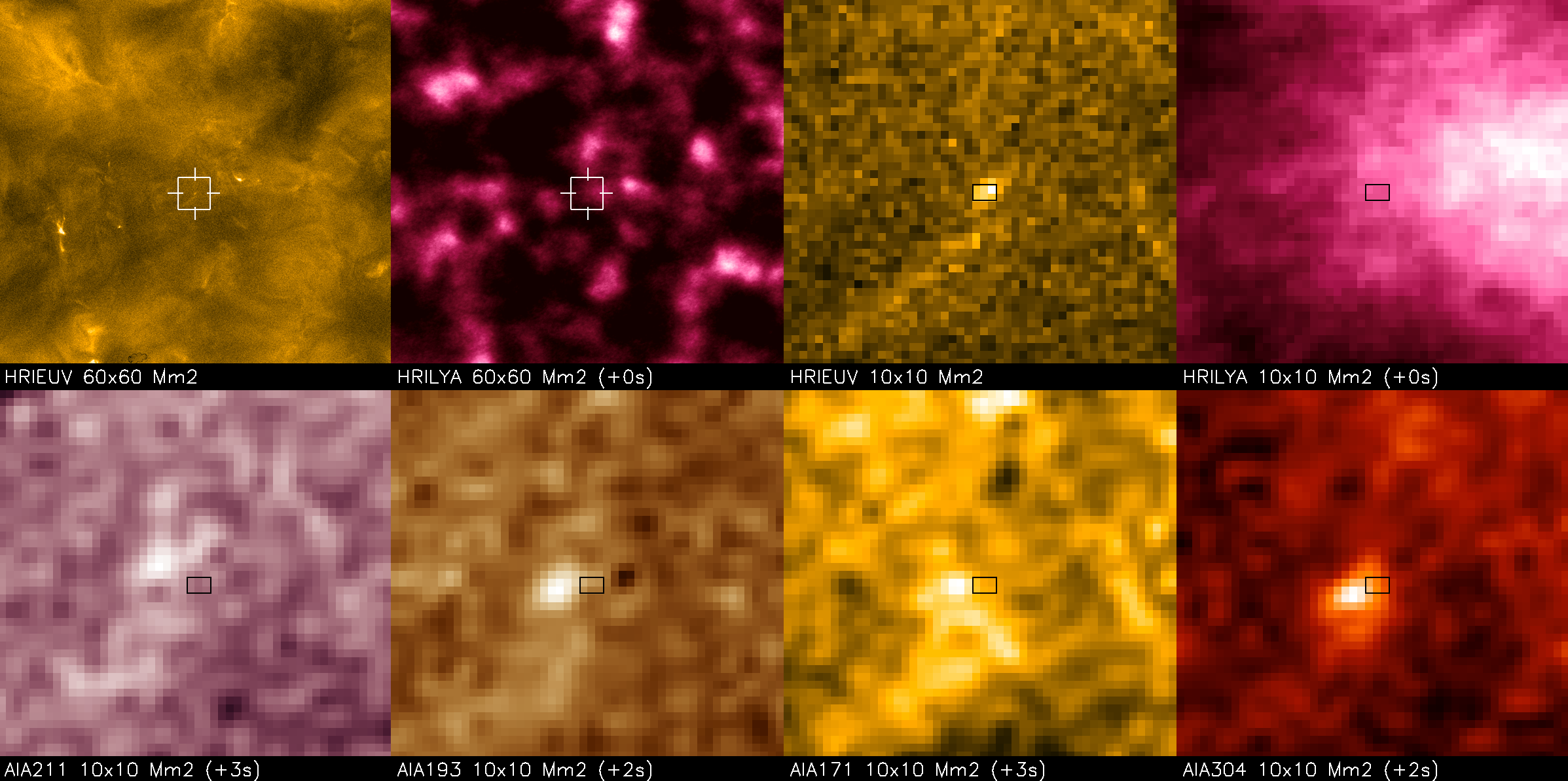}} 
\caption{Example of a dot-like event. The white square in the two top-left panels corresponds to the field of view of the other, zoomed-in panels.
}
\label{fig:cf104_BP5}
\end{figure*}

As expected from the histograms in Fig.~\ref{fig:histograms}, most campfires are very close to the \hrieuv resolution limit. In this section we do not consider events detected as a single pixel or appearing in a single frame. Even then, most campfires appear as essentially dot-like of only a few pixels but their existence is beyond doubt as we can see their evolution over several images. In Fig.~\ref{fig:cf709_BP2}, Fig.~\ref{fig:cf104_BP5}, and Fig.~\ref{fig:cf344_BP6}, we show examples of such dot-like campfires.  In each of the cases, there is no corresponding localised brightening in \hrilya. Comparison with simultaneous AIA images shows that these smallest events can be identified in the displayed AIA wavelengths also although they appear weaker, fuzzier and typically bigger which is most likely an effect of the lower spatial resolution. Without the indication from \hrieuv, most of these events would remain unnoticed among noise fluctuations.


\begin{figure*}[h]
\centering
{ \includegraphics[width=1.0\hsize]{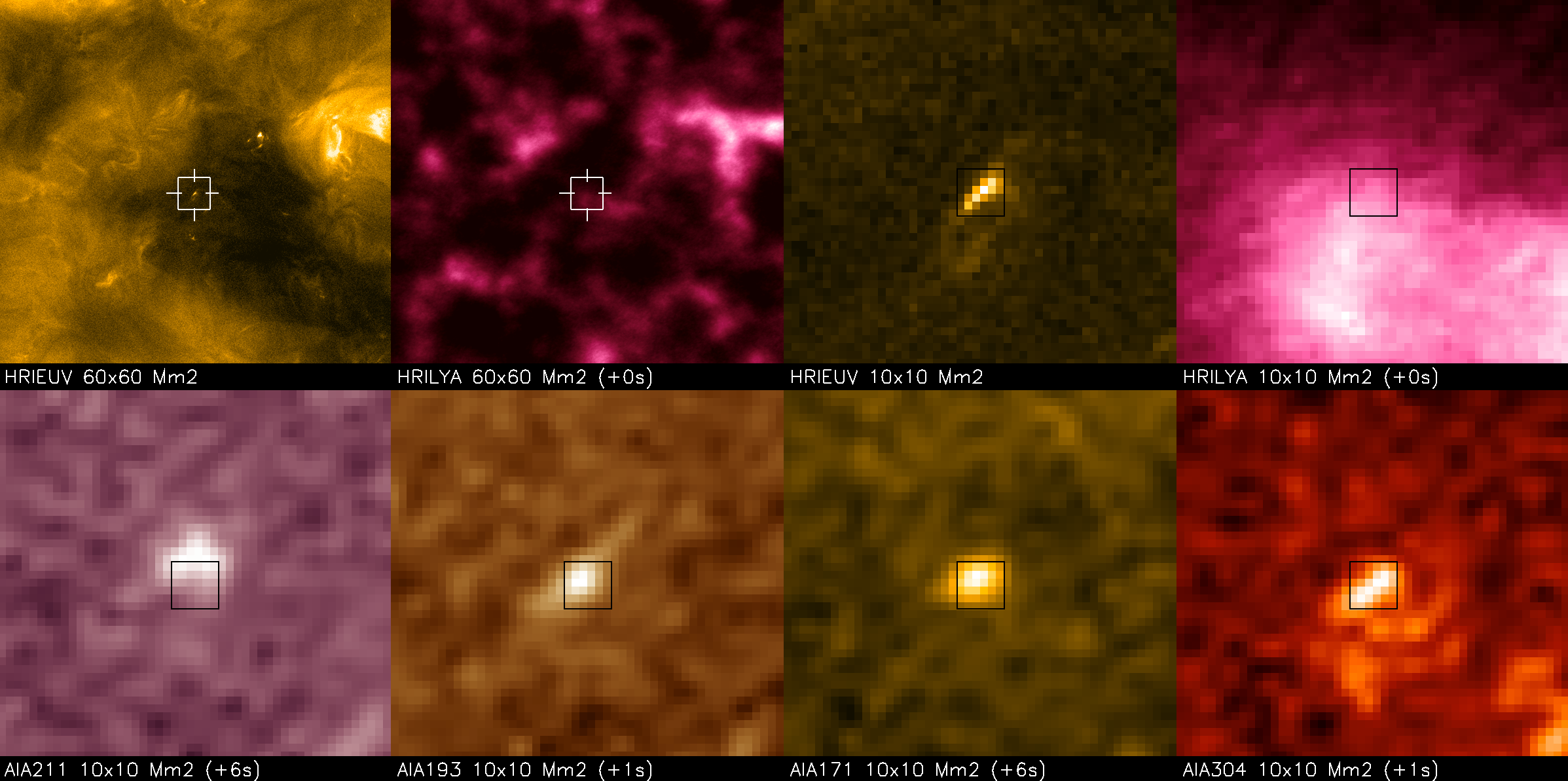}} 
\caption{Example of a loop-like event. The white square in the two top-left panels corresponds to the field of view of the other, zoomed-in panels.
}
\label{fig:cf8_BP10}
\end{figure*}

Amongst the slightly bigger campfires (e.g. Fig.~\ref{fig:cf8_BP10} and Fig.~\ref{fig:cf297_BP12}), we see loop-like events with an aspect ratio of 2 to 4 and a length of up to 10 pixels (2 Mm). Again,  there is no corresponding localised brightening in \hrilya but
the events can be identified in simultaneous AIA images, though they appear weaker and fuzzier, thereby often hiding their elongated structure.

\begin{figure*}[h]
\centering
{ \includegraphics[width=1.0\hsize]{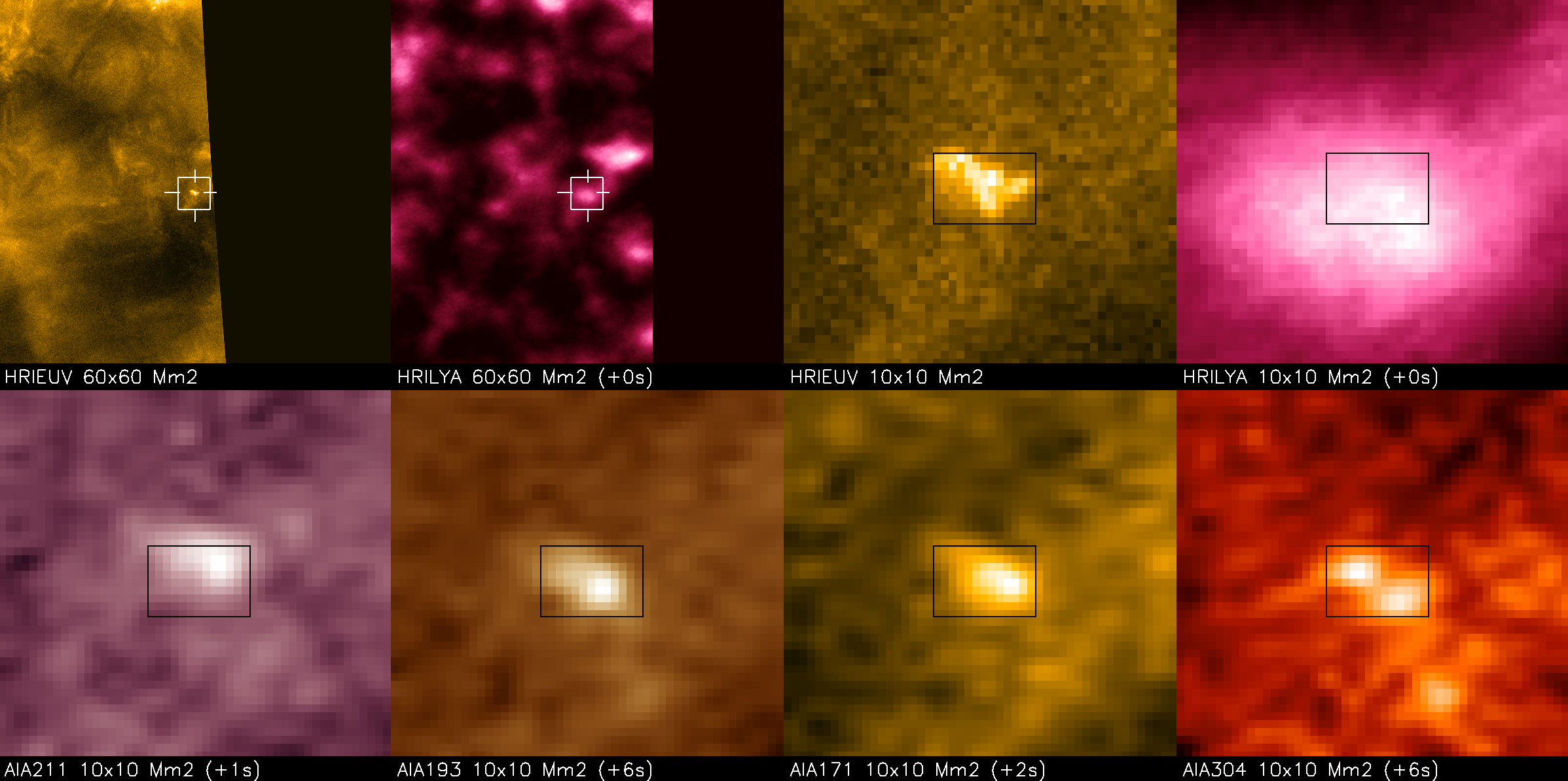}} 
\caption{Example of complex event. The white square in the two top-left panels corresponds to the field of view of the other, zoomed-in panels.
}
\label{fig:cf144_BP14}
\label{fig:FigC4}
\end{figure*}

\begin{figure*}[h]
\centering
{ \includegraphics[width=1.0\hsize]{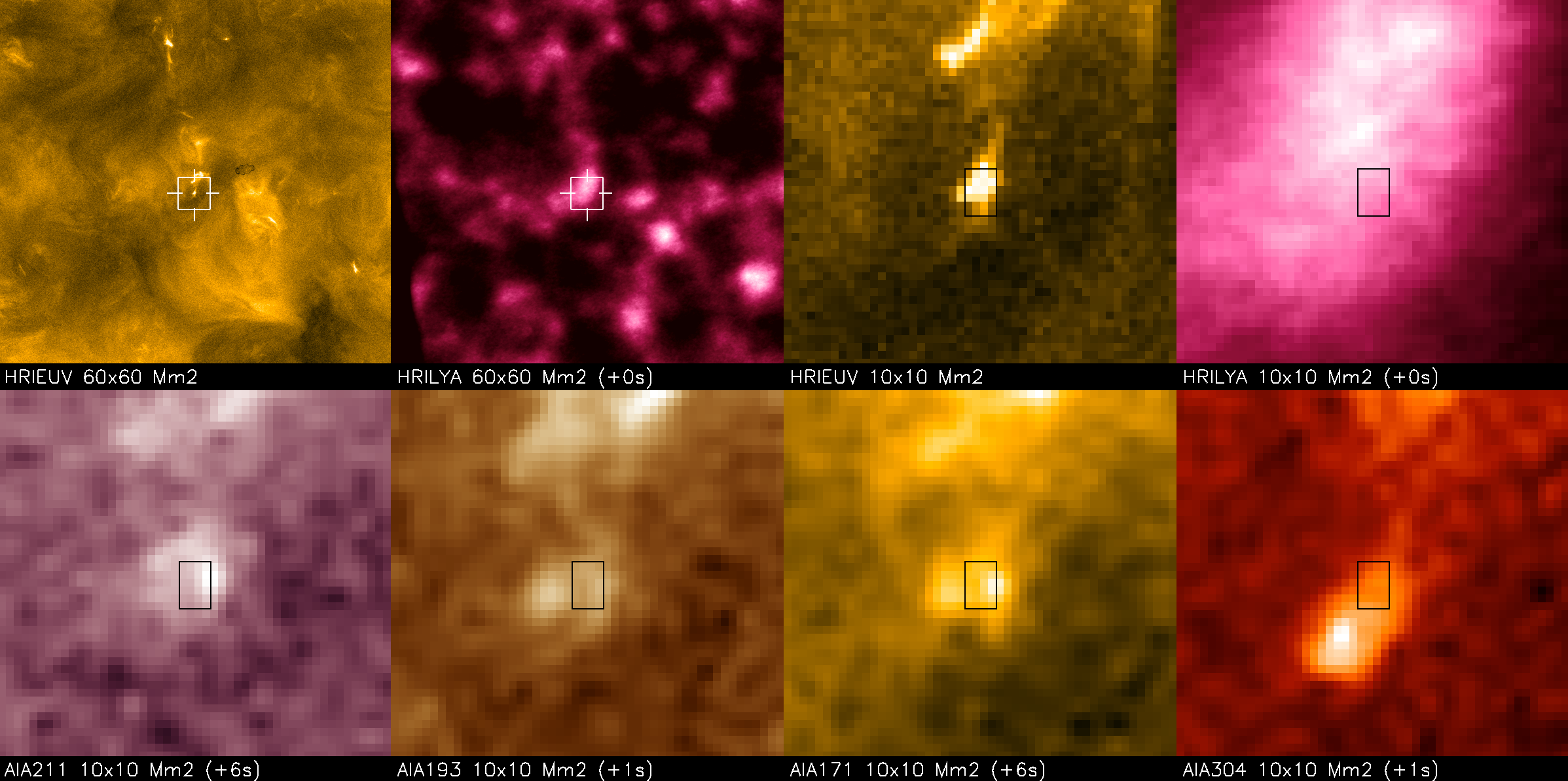}} 
\caption{Example of complex event. The white square in the two top-left panels corresponds to the field of view of the other, zoomed-in panels.
}
\label{fig:cf161_BP16}
\label{fig:FigC5}
\end{figure*}

\begin{figure*}[h]
\centering
{ \includegraphics[width=1.0\hsize]{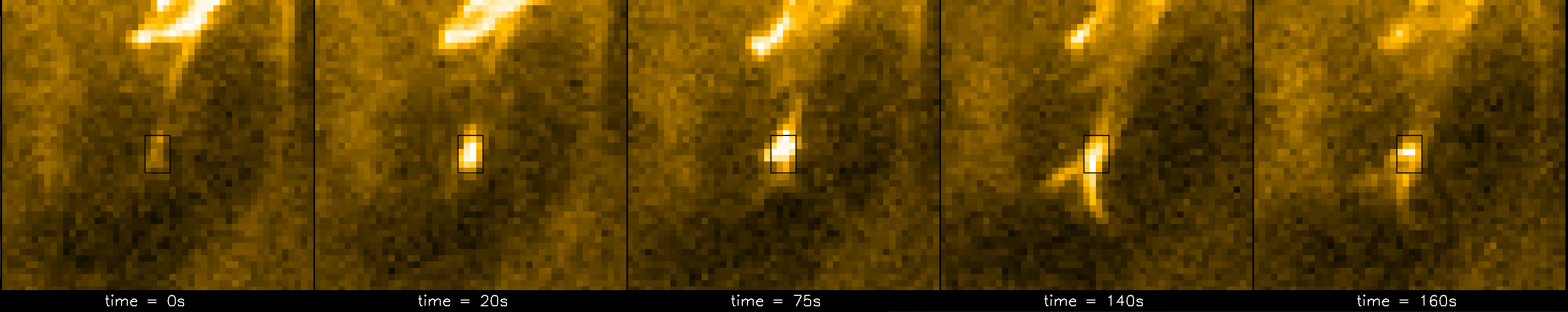}} 
\caption{Time evolution of the event shown in Fig.~\ref{fig:cf161_BP16}. The latter corresponds to t = 75s. Each frame corresponds to an area of \SI{10x10}{\mega\metre} and each pixel corresponds to an area of  \SI{197x197}{\kilo\metre}
}
\label{fig:InteractingLoops}
\end{figure*}

For the biggest campfires,  \hrieuv often shows complex internal structure, which is missing from the corresponding AIA images due to the difference in resolution (Fig.~\ref{fig:FigC4}). This internal structure can possibly be interpreted as a group of nearby loops brightening up, either because of some local source in their neighbourhood, or because of mutual interactions. A suggestive example of the latter is the event shown  in Fig.~\ref{fig:FigC5}. In the evolution of this event (Fig.~\ref{fig:InteractingLoops}) one can see that the source location of the event (the black rectangle) later becomes the forking point of what could be interpreted as two interacting loops.

\section{Height of campfires}
\label{section:Height of campfires}

In order to determine the height of campfires above the photospheric surface, we performed triangulation on sixteen small-size campfires that could be identified on both \hrieuv (14:54:10~UT) \SI{17.4}{\nano\metre} and AIA \SI{17.1}{\nano\metre} (14:57:57~UT) images.
The difference in time of the two images is chosen to accommodate for the light travel time to the two instruments, so that they image the same solar scene nearly simultaneously (to within 2~s).

The method consists of identifying the same feature (pixel) in the two images (tie-pointing) and calculating the lines of sight that pass through the corresponding pixels in the two images. This allows calculating the position of the intersection point in 3D space \citep[e.g.,][]{Inhester2006}. The triangulation was performed using the scc\_measure.pro SolarSoft routine, and the location of the campfires in Stonyhurst coordinates (longitude, latitude and height $h$ from the Sun center, see \citeauthor{Thompson2006} \citeyear{Thompson2006}) was obtained. The half-width triangulation error $\delta h$ is related to the linear pixel sizes $\delta s_1$ and $\delta s_2$ of \hrieuv  and AIA, respectively, and can be calculated as follows:
\begin{equation}
\delta h = \frac{\sqrt{\delta s_1^2 + \delta s_2^2 + 2 \, \delta s_1 \, \delta s_2 \, \cos \gamma}} {2 \sin \gamma}
\end{equation}
where $\gamma$ is the separation angle between Solar Orbiter and SDO. On  2020 May 30, $\gamma = 31.5^\circ$, one AIA pixel is 441~km and one \hrieuv  pixel is 198~km. The error is then $\delta h = 600$~km if one can localize a campfire to within one pixel in both \hrieuv and AIA images. This is sometimes difficult to do, so that the error may be up to four \hrieuv  and two AIA pixels,  which gives an error of 1500~km. This is the random error linked to the limited resolution of the telescopes.

On top of the random error for the height, there is a systematic error due to an uncertainty of a few arcsec on the pointing of the \hrieuv telescope with respect to the center of the Sun. We estimated this uncertainty to lead to a systematic error of $\pm 900$~km on the derived heights.

\end{appendix} 

\end{document}